# A new spin gapless semiconductor: quaternary Heusler CoFeCrGa alloy

*Lakhan Bainsla,[1,#] A. I. Mallick,[1,#] M. Manivel* Raja,[2] *A. A. Coelho,[3] A. K. Nigam,[4] D. D. Johnson[5,6], Aftab Alam[1] and K. G. Suresh[*,1]*

[1]Department of Physics, Indian Institute of Technology Bombay, Mumbai 400076, India
[2]Defence Metallurgical Research Laboratory, Hyderabad 500058, India
[3]Instituto de Física "Gleb Wataghin", Universidade Estadual de Campinas-UNICAMP, SP 6165, Campinas 13 083 970, Sao Paulo, Brazil
[4]DCMPMS, Tata Institute of Fundamental Research, Mumbai 4000005, India
[5]The Ames Laboratory, U.S. Department of Energy, Ames, Iowa 50011-3020, USA
[6]Department of Materials Science & Engineering, Iowa State UNiversity, Ames, Iowa 50011, USA

[#]L. Bainsla and A. I. Mallick contributed equally to this work.

## Abstract

Despite a plethora of materials suggested for spintronic applications, a new class of materials has emerged, namely spin gapless semiconductors (SGS), that offers potentially more advantageous properties than existing ones. These magnetic semiconductors exhibit a finite band gap for one spin channel and a closed gap for the other. Here, supported by the first-principles, electronic-structure calculations, we report the first experimental evidence of SGS behavior in equiatomic quaternary CoFeCrGa, having a cubic Heusler ($L2_1$) structure but exhibiting chemical disorder ($DO_3$ structure). CoFeCrGa is found to transform from SGS to half-metallic phase under pressure, which is attributed to unique electronic-structure features. The saturation magnetization ($M_S$) obtained at 8 K agrees with the Slater-Pauling rule and the Curie temperature ($T_C$) is found to exceed 400 K. Carrier concentration (up to 250 K) and electrical conductivity are observed to be nearly temperature independent, prerequisites for SGS. The anomalous Hall coefficient is estimated to be 185 S/cm at 5 K. Considering the SGS properties and high $T_C$, this material appears to be promising for spintronic applications.

**I. Introduction**

Recent studies have reported an interesting new class of materials called spin gapless semiconductors (SGS) [1,2]. SGS exhibit an open band gap for one spin channel and closed one for the other, rendering them with unique properties of half-metals and semiconductors simultaneously. This new class may be very effective in bridging between semiconductors and half-metallic ferromagnets. Diluted magnetic semiconductors (DMS) are of interest due to predicted semiconductor spintronic properties in ferromagnetic semiconductors. However, the major drawback of DMS is the low Curie temperature ($T_C$), which can be overcome in spin gapless semiconductors. The $T_C$ is high for some recently identified SGS, such as CoFeMnSi ($T_C$ = 620 K) [3] and $Mn_2CoAl$ ($T_C$ = 720 K) [2].

The electronic structure of SGS is different from that of half-metallic materials. Schematically, when the top of the valence band and bottom of the conduction band touch the Fermi energy for the majority-spin channel but shows a semiconducting gap for the minority-spin channel, the resulting band structure will be of SGS. The schematic density of states (DOS) for half metals and spin gapless semiconductors are compared in Fig. 1. The unique electronic structure in SGS materials gives rise to some interesting properties that make them quite attractive from fundamental and applied point of view.

Very recently, promising SGS behavior in Heusler alloys [2,4] has been identified, where cubic Heusler phase has stoichiometry $X_2YZ$, where X and Y are the transition metals and Z is a nonmagnetic element. Xu et al. noted in their recent report [5] that Heusler alloys with 26 or 28 valance electrons have a great potential to exhibit SGS behavior. Quaternary compounds XX'YZ with the stoichiometry 1:1:1:1 show the LiMgPdSn prototype or the so-called Y-type structure (space group #216) with somewhat different symmetry [6]. To the best of our knowledge, quaternary Heuslers have been explored only a very little. Although high spin polarization is proposed in a few related Heusler alloys, such as CoFeMnZ (Z=Al, Ga, Ge), they show half-metallic (not SGS) behavior [6,7,8]. In addition, there is a lack of magneto-transport measurements in such compounds, which are considered to be a more careful probe for predicting the correct behavior of such magnetic semiconductors.

Our main motivation is to find a material candidate in the XX'YZ family, which shows a SGS behavior having a high Curie temperature. CoFeCrGa (CFCG) Heusler is one such compound that shows the desired properties. We have performed detailed structural, magnetic, resistivity and Hall measurements to establish the SGS behavior in this compound. To the best of our knowledge, this is the first time that CFCG is synthesized and characterized by any technique. Our *ab-initio* calculations on CFCG also supported it SGS behavior. The origin of the dispersion (band gap) near the Fermi energy in different spin directions is discussed. Because these compounds are sensitive to the external effects (e.g. pressure, magnetic field, and doping), we have also studied the effect of pressure. CFCG is found to transform from SGS to half-metallic behavior under pressure, with an increase in the minority-spin band gap. The calculated and experimental magnetic moment is found to follow the Slater-Pauling rule, which is a prerequisite criterion for SGS material.

## II. Experimental Details

*Sample preparation:* The polycrystalline bulk sample of CFCG was prepared by arc melting the appropriate quantities of various elements (at least 99.9 % purity). Ingot was flipped and melted several times for homogeneity. To further increase the homogeneity arc-melted samples were annealed at 1073 K for two weeks.

*Characterization:* The crystal-structural studies were carried out at room temperature using x-ray diffraction pattern obtained with Cu-K$_\alpha$ radiation and $^{57}$Fe Mössbauer spectra recorded using a constant acceleration spectrometer with 25 mCi $^{57}$Co(Rh) radioactive source. The obtained Mössbauer spectra were analyzed using PCMOS-II least-square fitting program. The phase stability of the alloy was checked by differential thermal analysis (DTA). The magnetic and transport measurements were performed in the temperature range of 5–300 K and in fields up to 50 kOe using the PPMS (Quantum Design). Magnetization measurements were performed under applied hydrostatic pressures. The carrier concentration, $n = 1/eR_H$ was obtained from $R_H$ (Hall coefficient) using the single-band model [9].

**III. Results and discussion**

The Rietveld refinement of room temperature x-ray diffraction pattern is shown in Fig. 2(a). The alloy exhibits cubic Heusler structure (prototype LiMgPdSn), with lattice parameter (*a*) of 5.79 Å. Interestingly, the superlattice reflections (111) and (200) are absent. When the main group element (here Ga) is from the same period of the system as the transition metals (here Co, Fe and Cr), it is very difficult to find out the correct structure unambiguously by the x-ray or neutron diffraction data. Superlattice reflections were not observed in the XRD patterns of $Co_2FeZ$ (Z=Al, Si, Ga, Ge) alloys, but the extended x-ray absorption fine structure technique showed the existence of $L2_1$ order [10]. Importantly, a well-ordered crystal structure is a necessary requirement for high spin polarization of the materials because the tunneling magnetoresistance (TMR) ratio is related to the spin polarization [11].

To further investigate the crystal structure, we have performed $^{57}$Fe Mössbauer spectroscopic measurements at room temperature, as shown in Fig. 2(b). The best fit to the data was obtained with two sextets ($S_1$ and $S_2$) having hyperfine field ($H_{hf}$) values of 254 ($S_1$) and 142 ($S_2$) kOe and relative intensities of 58 ($S_1$), 34 ($S_2$), respectively, along with a doublet 8% (paramagnetic). The small value of quadrupole shift is in agreement with the cubic symmetry of the local environment of Fe. In an ordered $L2_1$ structure, Fe atoms must occupy the Y sites with the cubic symmetry ($O_h$), which will result in a single sextet due to the presence of only one crystallographic site for Fe. The presence of second sextet could be attributed to the occupancy of Fe atoms at X, X' or Z sites, which indicates some amount of structural disorder. The hyperfine field values are not expected to change much when Fe occupies Z site because of the same number of magnetic near neighbors for Y and Z sites. On the other hand, when Fe occupies X and X' sites, a large decrease in $H_{hf}$ is expected as X site has the highest number of non-magnetic near neighbors. The experimentally observed values of $H_{hf}$ (254, 142 kOe) clearly indicate that the Fe also occupies X, X' sites, resulting in the $DO_3$-type disorder. The sub-spectra $S_1$ and $S_2$ are ascribed to ordered $L2_1$ and $DO_3$ phases with relative intensities of 58% and 34%, respectively, implying that the structure is reasonably ordered at room temperature. The phase stability of the crystal structure was investigated by performing the DTA in the temperature range of 500 - 1600 K, as shown in the inset of Fig. 2(a). A peak in the DTA curves is obtained around 1000 K, which may be attributed to a $L2_1$–to–B2 structural transition.

The isothermal magnetization curves obtained under applied pressure (*P*) values at 8 K are shown in Fig. 3. The $M_S$ value of ≈ 2.1 $\mu_B$/f.u. is found at 8 K under zero applied pressure, in good agreement with the $M_S$ value (2.0 $\mu_B$) from the Slater-Pauling rule [12]. This value is found to be constant with the application of pressure, and is in agreement with our theoretical results (see below), as well as and those by Picozzi et al. for half-metallic ferromagnets [13]. The pressure-independent $M_S$ values we found also for half-metallic $Co_2TiGa$ Heusler [14]. The Curie temperature of CFCG is expected to be above 400 K as no magnetic transition is observed up to that temperature during measurments.

The temperature dependencies of electrical conductivity ($\sigma_{xx}$) with zero applied field and carrier concentration [*n(T)*] are shown in Fig. 4. With increasing temperature $\sigma_{xx}$ increases and shows a non-metallic conduction behavior. The value of $\sigma_{xx}$ is found to be 3233

S/cm at 300 K, which is slightly higher than that obtained for other SGS Heusler alloys, e.g. $Mn_2CoAl$ = 2440 S/cm [2]. The $\sigma_{xx}$-$T$ behavior is unusual and is different from that of normal metals or semiconductors. The carrier concentration was determined from the Hall coefficient measurements in the range of 5 - 280 K. $n(T)$ is of the order of $10^{20}$ cm$^{-3}$ and almost temperature independent up to 250 K, above which it increases abruptly. The physical reasons behind the nearly temperature-independent carrier concentration in gapless semiconductors, as compared to the exponential dependence in the case of conventional semiconductors, are well known [15,16]. The $n(T)$ values obtained here (up to 250 K) is above that in HgCdTe ($10^{15}$ – $10^{17}$ cm$^{-3}$) [15] and $Mn_2CoAl$ ($10^{17}$ cm$^{-3}$) [2] but below that of $Fe_2VAl$ ($10^{21}$ cm$^{-3}$) [17], a proposed semiconducting Heusler. The abrupt change in $n(T)$ above 250 K may be due to the onset of thermal excitations across the half metallic band gap. The $n(T)$ and $\sigma_{xx}$ behavior is clear evidene for SGS behavior in this material.

The anomalous Hall conductivity $\sigma_{xy} = \rho_{xy}/\rho_{xx}^2$ at 5 K was obtained from the magnetic-field-dependent transport measurements, see Fig. 5. $\sigma_{xy}$ obtained for CFCG is comparable to those obtained for other half-metallic Heuslers. Its behavior is identical to the magnetization isotherms (see inset). The anomalous Hall conductivity ($\sigma_{xy0}$) is calculated as the difference in $\sigma_{xy}$ values at zero and the saturation fields. $\sigma_{xy0}$ attains a value of 185 S/cm, which is higher than that observed in $Mn_2CoAl$ (22 S/cm) [2], but less than that of the half-metallic $Co_2FeSi$ ($\approx$ 200 S/cm at 300 K) [18] and $Co_2MnAl$ ($\approx$ 2000 S/cm) [19].

**IV. Electronic structure calculations**

To further investigate the electronic properties of CFCG Heuslers, we have performed first-principles, electronic-structure calculations using spin-polarized density functional theory, as employed in Vienna *ab-initio* simulation package (VASP) [20] based on a projected-augmented wave basis [21]. The exchange-correlation functional was based on the generalized gradient approximation (GGA). . A 16x16x16 Monkhorst-pack k-point mesh was used for the Brillouin zone integration. We have used a plane-wave cut-off of 340 eV with the convergence criteria of 0.1 meV/cell (10 kbar) for energy (stress).

The prototype of a quaternary Heusler with the composition XX'YZ and space group F-43m (#216) is LiMgPdSn. There are four non-equivalent configurations based on the occupation of various Wyckoff sites by different constituent elements. From the total energy calculations we found the configuration with Co at X(0,0,0), Fe at X'(1/2,1/2,1/2), Cr at Y(1/4,1/4,1/4), and Ga at Z(3/4,3/4,3/4) to be the most stable.

The spin-resolved dispersion and density of states (DOS) of CFCG in the most stable configuration with experimental lattice parameter ($a_{\text{expt}}$ = 5.79 Å) are shown in Fig. 6. A closed band-gap character in the majority-spin state and a small open band gap (near the Fermi energy, $E_F$) in the minority-spin state suggest CFCG to behave as a spin gapless semiconductor. The valence band maximum for minority-spin state is slightly above $E_F$, yielding a negligibly small DOS at $E_F$, which arises from mixed contributions of *d*-bands from Co, Fe and Cr. A careful analysis of various bands crossing the $E_F$ for the minority-spin state (right panel) indicates that such small DOS at $E_F$ arises mainly from three bands, two of which are degenerate and are composed of ~54% Co and ~46% Fe $e_g$ sub-band characters (i.e., $d_{x2-y2}$ and $d_{3z2-1}$). The third band is contributed almost equally by $t_{2g}$ sub-bands (i.e., $d_{xy}$, $d_{yz}$, and $d_{xz}$) of Co (35%), Fe (29%) and Cr (32%).

To study the behavior of CFCG under pressure, we have calculated DOS in the most stable configuration versus lattice parameters below $a_{\text{expt}}$, which are plotted in Fig. 7. Notably, the behavior of CFCG changes from SGS to half-metallic with the decrease of lattice parameter. In the majority-spin state, the DOS at $E_F$ increases significantly from almost zero value at $a_{\text{expt}}$ to a finite value (under pressure), see Fig. 8(a). In addition, the band gap ($\Delta E_g$) at $E_F$ in the minority-spin state increases with lattice parameters decreasing below $a_{expt}$, indicating the collective effect of transition. The majority DOS at $E_F$ and $\Delta E_g$ with varying lattice parameters are shown in Fig. 8(a).

Figure 8(b) shows the variation of total magnetic moment per formula unit and the individual atomic magnetic moments (on Co, Fe and Cr atoms) of CFCG versus decreasing lattice parameter (*a*). The Fe moment is antiferromagnetically aligned with respect to Co and Cr for all *a*. The total moment remains constant with the decrease of *a*, supporting our magnetization measurements under pressure (see Fig. 3). Although the Co-moment remains unaffected, hydrostatic pressure lowers the Cr-moment and enhances the Fe-moment, collectively resulting in a constant total magnetic moment. The calculated (experimental) total magnetic moment $\mu_{tot}$ ~1.98 $\mu_B$ (~2.07 $\mu_B$) agrees well with the linear Slater-Pauling curve given by $\mu_{tot} = Z_t - 24$, where $Z_t$ is the total valence electrons per formula unit (26 for CFCG). The agreement with the Slater-Pauling behavior is one of the prerequisites for SGS.

## V. Conclusions

In conclusion, a detailed characterization of the equiatomic quaternary CoFeCrGa Heusler alloy reveals very clear signatures of SGS behavior. Such a behavior is also confirmed by our first-principles electronic-structure results. Characterization of various properties of the material has been done using some of the most reliable probes, i.e., x-ray diffraction, Mössbauer spectroscopy, and magnetization and transport measurements. The CoFeCrGa cubic Heusler (prototype LiMgPdSn) exhibits some amount of disorder (yielding DO$_3$ symmetry), though DTA suggest a L2$_1$-to-B2 structural transition at about 1000 K. Pressure-independent saturation magnetization values support a half-metallic electronic structure. The measured electrical conductivity (~3x10$^3$ S/cm) and nearly temperature-independent carrier concentration (~10$^{20}$ cm$^{-3}$) up to 250 K suggest that the CoFeCrGa is a SGS up to 250 K. Under pressure, CFCG is found to transform from SGS to half-metallic beahvior. The saturation magnetization, supported by DFT calculations, follows the linear

Slater-Pauling rule, which is a prerequisite criterion for SGS materials. We plan to investigate whether the SGS electronic structure at room temperature can be retained via appropriate alloy substitutions in the parent CFCG Heusler. Considering the high $T_C$ and proposed SGS behavior, this material may a potential candidate for semiconductor spintronic devices, such as spin injection.

**Acknowledgements**

One of the authors, LB, would like to thank UGC, Government of India for granting senior research fellowship (SRF). The authors convey their thanks to D. Buddhikot for his help in the resistivity measurements. KGS thanks ISRO, Govt. of India for the financial support. AIM acknowledges the support from TAP fellowship under SEED Grant (project code 13IRCCSG020). Support at Ames Laboratory (DDJ) was from the Department of Energy, Office of Science, Basic Energy Science, Materials Sciences and Engineering Division. The Ames Laboratory is operated for the U.S. DOE by Iowa State University under Contract No. DE-AC02-07CH11358. We acknowledge support for computing time from Ames Laboratory.

**Figures**

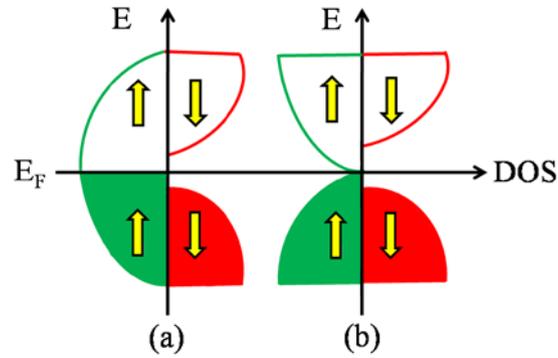

FIG. 1. (Color online) Density of states scheme for a typical (a) half metal and (b) spin gapless semiconductor.

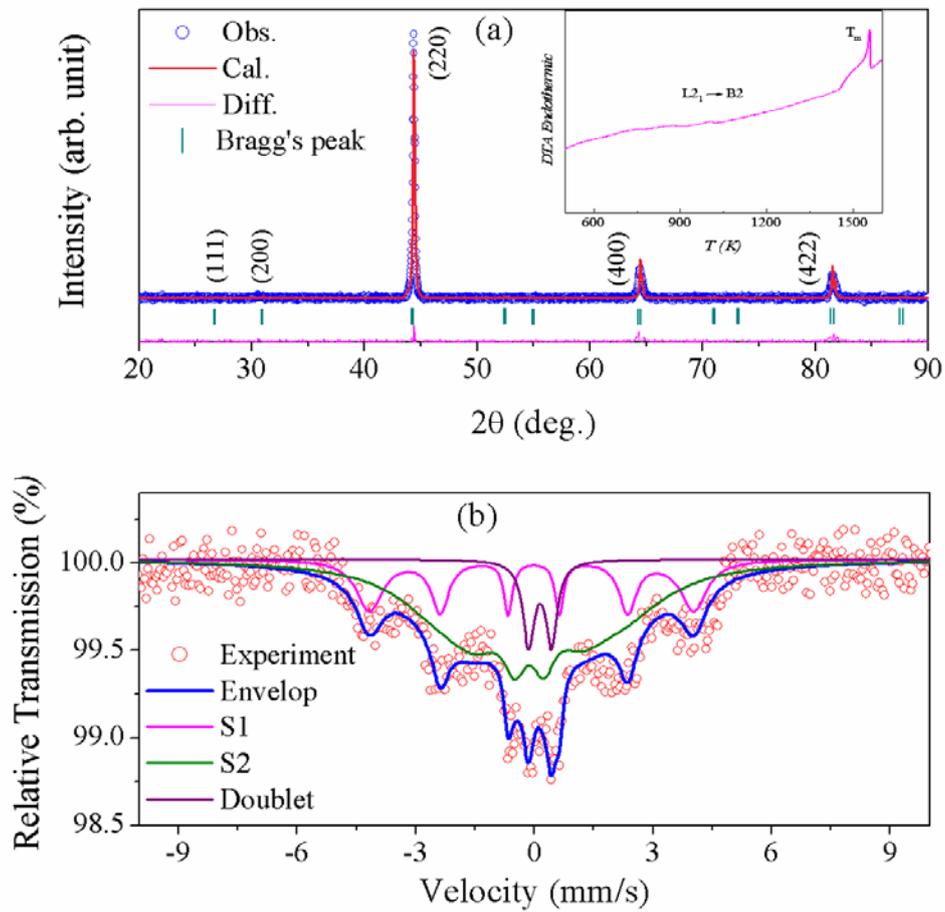

FIG. 2. (Color online) (a) Rietveld refinement (300 K) x-ray diffraction pattern of CFCG collected using Cu-K$_\alpha$ radiation. (Inset) DTA curves obtained from 500 - 1550 K. (b) $^{57}$Fe Mössbauer spectra of CFCG (300 K).

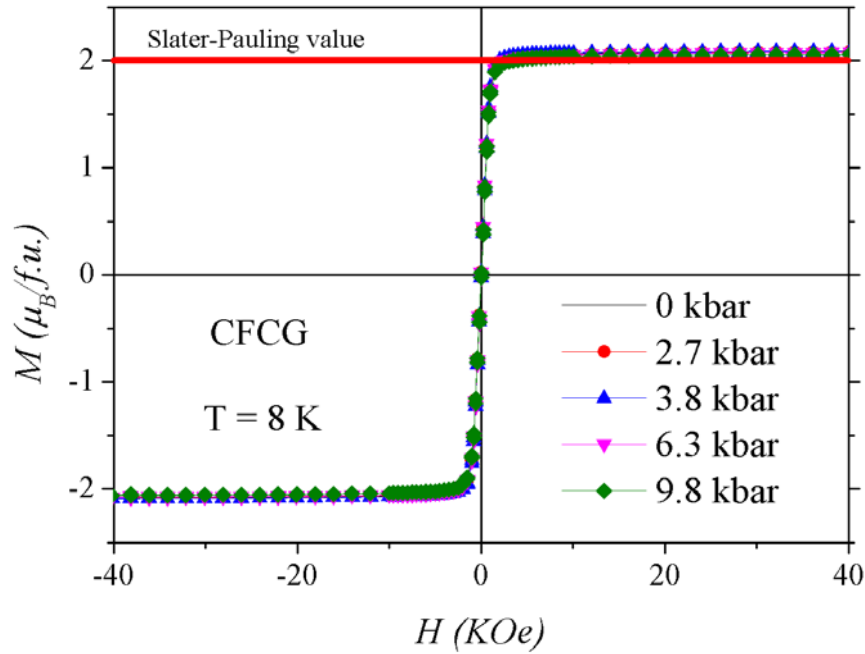

FIG. 3. (Color online) CFCG isothermal magnetization curves at 8 K and at various pressures.

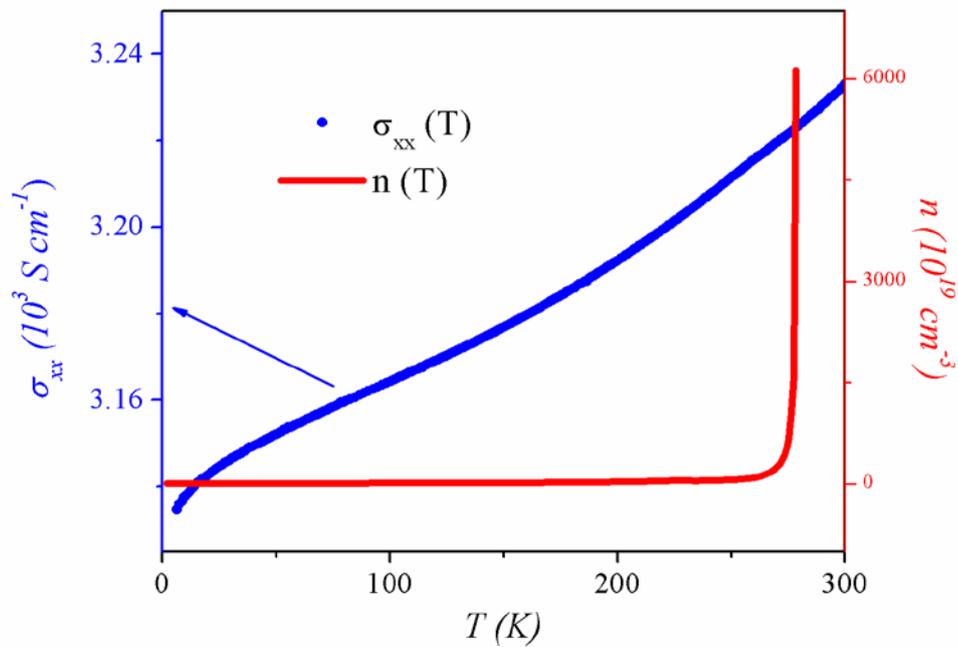

FIG. 4. (Color online) (Left) Electrical conductivity ($\sigma_{xx}$) and [right] carrier concentration $n(T)$ from 5 - 300 K of CFCG.

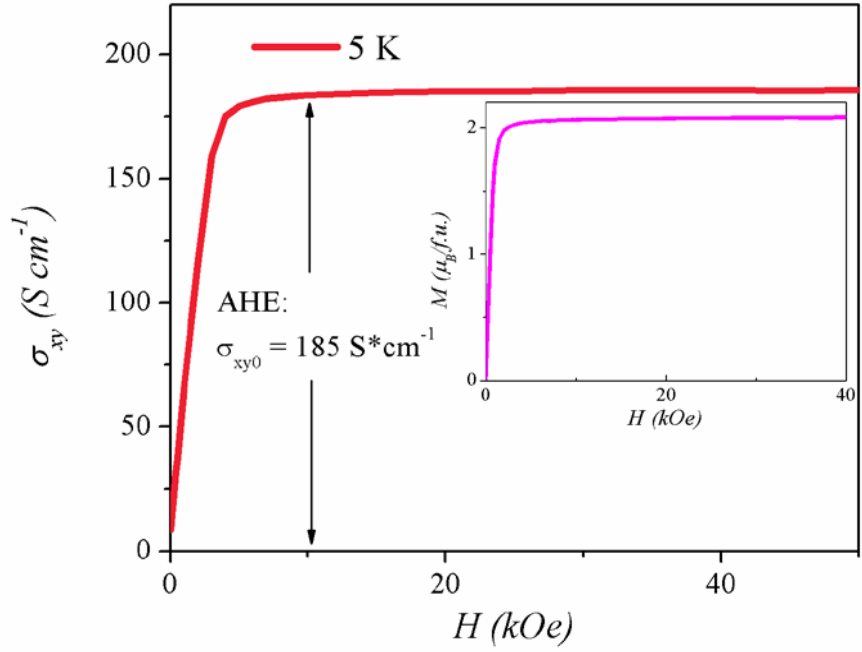

FIG. 5. (Color online) Hall conductivity ($\sigma_{xy}$) of CFCG versus applied field. The inset shows the magnetization isotherm at 8 K.

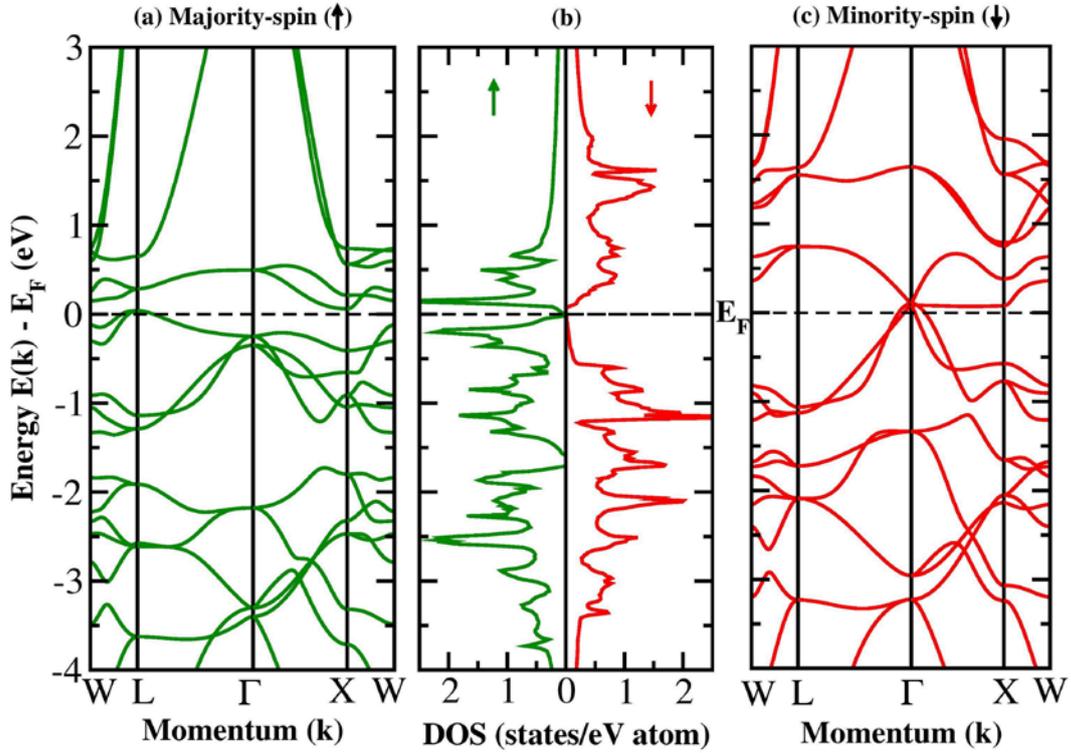

FIG. 6. (Color online) Dispersion and density of states of CFCG: (a) majority-spin states, (b) density of states, (c) minority-spin states at $a_{expt}$ (5.79 Å).

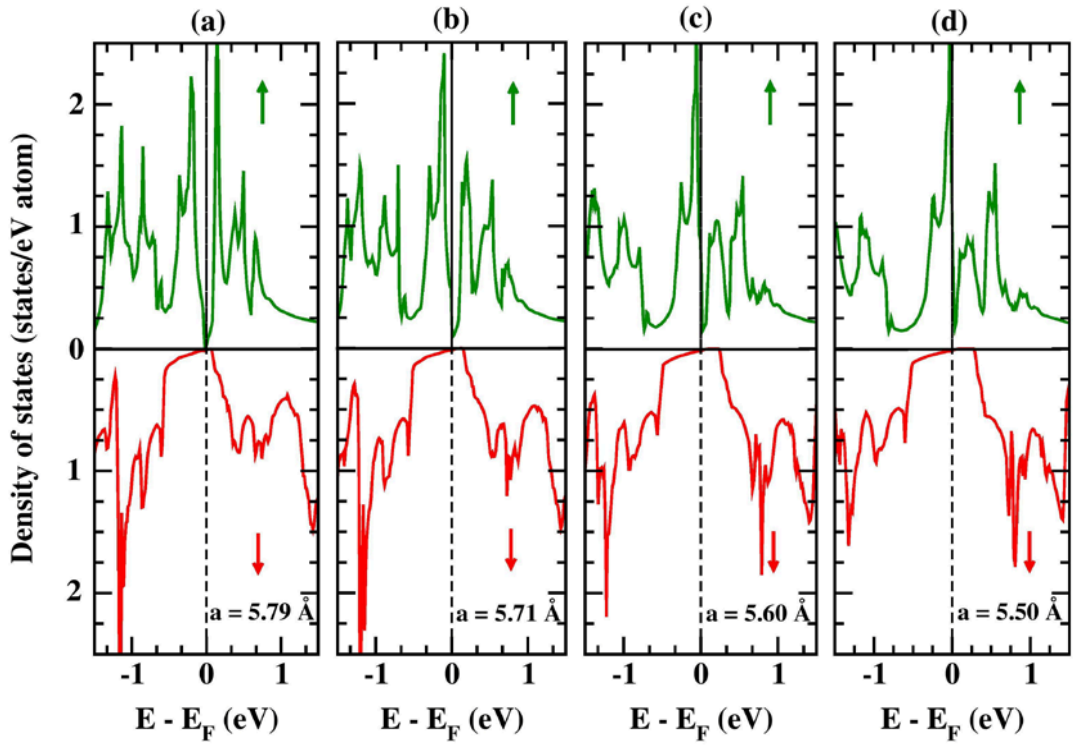

FIG. 7. (Color online) CFCG DOS versus $a$ (Å): (a) 5.79 ($a_{expt}$), (b) 5.71, (c) 5.60, and (d) 5.50.

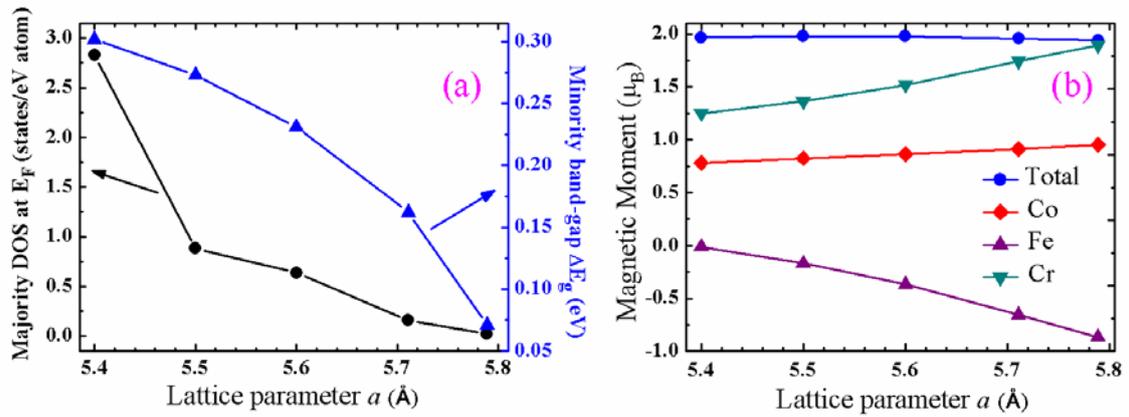

FIG. 8. (Color online) For CFCG, versus $a$ the (a) DOS at $E_F$ in the majority-spin state (left-hand scale) and band-gap in the minority-spin state (right-hand scale) and (b) total and site magnetic moments.